\documentclass[conference]{IEEEtran}
\IEEEoverridecommandlockouts

\usepackage{amsmath,amssymb,amsfonts}
\usepackage{graphicx}
\usepackage{textcomp}
\usepackage{xcolor}
\usepackage{caption}
\usepackage{subcaption}
\usepackage{etoolbox}
\usepackage{float}
\usepackage{url}
\usepackage{xurl}
\usepackage{stfloats}
\usepackage{comment}
\usepackage{hyperref}
\usepackage[ruled,vlined]{algorithm2e}
\usepackage{enumitem}
\usepackage{makecell}
\UseRawInputEncoding

\BeforeBeginEnvironment{appendices}{\clearpage}

\usepackage{listings}
\usepackage{xcolor}

\definecolor{codegreen}{rgb}{0,0.6,0}
\definecolor{codegray}{rgb}{0.5,0.5,0.5}
\definecolor{codepurple}{rgb}{0.58,0,0.82}
\definecolor{backcolour}{rgb}{0.95,0.95,0.92}

\lstdefinestyle{mystyle}{
    backgroundcolor=\color{backcolour},   
    commentstyle=\color{codegreen},
    keywordstyle=\color{magenta},
    numberstyle=\tiny\color{codegray},
    stringstyle=\color{codepurple},
    basicstyle=\ttfamily\footnotesize,
    breakatwhitespace=false,         
    breaklines=true,                 
    captionpos=b,                    
    keepspaces=true,                 
    numbers=left,                    
    numbersep=5pt,                  
    showspaces=false,                
    showstringspaces=false,
    showtabs=false,                  
    tabsize=2
}

\usepackage{cite}
\def\BibTeX{{\rm B\kern-.05em{\sc i\kern-.025em b}\kern-.08em
    T\kern-.1667em\lower.7ex\hbox{E}\kern-.125emX}}
\usepackage{pifont} 

\newcommand{\cmark}{\ding{51}} 
\newcommand{\xmark}{\ding{55}} 

\begin{document}

\title{Explain First, Trust Later: LLM-Augmented Explanations for Graph-Based Crypto Anomaly Detection}

\author{
    \IEEEauthorblockN{Adriana Watson, M.S.}
    \IEEEauthorblockA{\textit{School of Engineering Technology} \\
    \textit{Purdue University}\\
    West Lafayette, United States \\
    watso213@purdue.edu}
    \and
    \IEEEauthorblockN{Grant Richards, Ph.D.}
    \IEEEauthorblockA{\textit{School of Engineering Technology} \\
    \textit{Purdue University}\\
    West Lafayette, United States \\
    grichard@purdue.edu}
    \and
    \IEEEauthorblockN{Daniel Schiff, Ph.D.}
    \IEEEauthorblockA{\textit{College of Liberal Arts} \\
    \textit{Purdue University}\\
    West Lafayette, United States \\
    dschiff@purdue.edu}
}

\maketitle

\begin{abstract}
The rapid growth of decentralized finance (DeFi) has been paralleled by an alarming rise in cryptocurrency-related financial crime \cite{federal_bureau_of_investigation_cryptocurrency_2023}. Traditional fraud detection systems struggle in this domain due to the sparsity of anomalies in the dataset, the black-box nature of modern machine learning models, and the absence of human-intelligible explanations \cite{wessan_problems_2024}. This work introduces a modular pipeline that (i) employs a graph neural network (GNN) to score wallet-level anomalies, (ii) applies GraphLIME to produce local, feature-level attributions, and (iii) uses a large language model (LLM) to translate attributions and node statistics into concise, natural language narratives for human-in-the-loop review. Using a real-world, graph-structured Bitcoin dataset (Elliptic++) \cite{elmougy_demystifying_2023}, the approach flags suspicious wallets, surfaces the features most responsible for each decision, and generates calibrated explanations that explicitly separate importance scores from observed values. The design is regulation-agnostic and model-agnostic, enabling substitution of different GNNs, explainers, or LLMs, and includes a lightweight dashboard to support investigator triage. 
\end{abstract}


\section{Introduction}
As the popularity of cryptocurrency and other decentralized finance mediums has grown, so have concerns surrounding the rising trend of cryptocurrency attacks such as money laundering, phishing, and Ponzi schemes. These attacks are constantly evolving, rendering existing fraud detection systems unreliable. With the threat of cryptocurrency attacks becoming increasingly minacious, the need for novel solutions to identify cryptocurrency fraud has become apparent. 

The emergence of machine learning technology has promised a new wave of solutions to conventionally complex tasks, including fraud detection. With ever-evolving banking methods and increasingly clever attacks, the need for more sophisticated countermeasures is clear. 

Machine learning offers strong capabilities for complex detection tasks in financial domains, yet practical deployment is limited by model opacity and severe class imbalance as illicit activity is rare. Existing cryptocurrency and banking tools, including large bank systems and credit card oriented detectors, reflect these challenges \cite{jp_morgan_documentation_nodate,bin_sulaiman_review_2022}. Performance and usability are often reduced by black box decision making and the difficulty of learning from sparse anomalies \cite{chaquet-ulldemolins_black-box_2022,elmougy_demystifying_2023,wessan_problems_2024}.

To address these constraints, this work introduces a multistage, regulation-agnostic pipeline that integrates graph based anomaly detection with explanation and human oversight. A graph neural network scores wallet level anomalies. GraphLIME provides local, feature level attributions for flagged instances. A large language model converts GraphLIME feature importance weights and node statistics into concise narratives suitable for audit and regulatory review, and a human-in-the-loop audit layer reduces false positives and supports accountable decision making.

The contributions are fourfold. First, a transparent, graph-first pipeline is formulated that joins a graph neural network for anomaly scoring with GraphLIME for local attributions and a large language model for case narratives. Second, prompting strategies are specified that separate feature importance from observed values in order to avoid conflation in explanations. Third, an evaluation protocol is outlined that spans anomaly metrics, explanation faithfulness, and analyst utility. Fourth, a lightweight dashboard is provided that fuses graph context, attributions, and narratives to support investigator triage.

\section{Literature Review}
Existing work documents a widening range of cryptocurrency abuses, including laundering, phishing, and Ponzi schemes, which complicate oversight and enforcement because the fraudulent action can be hard to localize and identify \cite{trozze_cryptocurrencies_2022,leuprecht_virtual_2022}. As cryptocurrency systems mature, many deployed fraud controls remain poorly matched to the relational and dynamic nature of decentralized finance.

\subsection{Motivation} \label{sec: motivation}

The challenge of catching and prosecuting cryptocurrency crimes is ongoing and highlights a clear divide between the offerings of existing solutions and actual applications. As a result, cryptocurrency fraud attacks have devastating and unresolved impacts on their victims. 

Yet, companies investigated by the Market Integrity and Major Frauds Unit (MIMF) under the Department of Justice Criminal Division, were most commonly prosecuted for violations of lesser crimes than the ones committed likely because identifying the actual incursion would be too difficult. For example, BitConnect, a company that carried out a \$2.4 billion Ponzi scheme was ultimately charged with wire fraud, operating an unlicensed money transmitting business, and conspiracy \cite{noauthor_united_2021}. Despite the \$2.4 billion lost by users, only \$17 million was paid out to the victims of the company, likely in part due to these insufficient charges \cite{internal_revenue_service_victims_2023}. Similarly, Forsage, a company that defrauded investors out of \$340 million was only charged with two counts of conspiracy to commit wire fraud \cite{noauthor_united_2023}. At the time of writing, the victims of the scheme have not been compensated. 

A similar theme plays out in cases brought by the U.S. Securities and Exchange Commission (SEC). Of the cases presented by the SEC that were actually prosecuted for fraud, the most common charge was under securities fraud statutes \cite{us_securities_and_exchange_commision_crypto_2025}. While the definition of this charge (a misrepresentation of securities offerings) fits the crime to an extent, Ponzi schemes, rug-pulls, and other cryptocurrency attacks are much more devastating \cite{kutera_cryptocurrencies_2022}. Given the glaring lack of legislation to adequately mete out justice for these types of violations, key interim solutions will likely be found by the more nimble private sector. Moreover, emerging academic research may hold the key to bridging the gap between the profit-driven private sector and technologically lagging public sector. Such a link is critical for both advancing cryptocurrency fraud detection and informing legislation. For this reason, a cryptocurrency fraud detection solution that enables regulatory enforcement can only be effective if it has the following traits:
\begin{enumerate}
    \item The fraud detection model must be trained on a graphical database so transaction and wallet metadata is maintained. 
    \item If the solution uses a black-box model, it must have an explainable component. This serves as evidence for how a particular decision was made by the model which is essential for any application with real-world implications (such as possible prosecution). 
    \item The output of the solution must be coherent for a non-technical audience. While the final user may be a subject matter expert, it can not be assumed that all players are at the same level as fraud investigation requires a bread range of players. 
\end{enumerate}
These features combined would produce a solution that can more effectively identify financial crime in the sea of transaction data that exists without losing key context for suspicious events. This allows regulatory bodies to catch these financial crimes more quickly with a lower risk of false positives and much less manual investigation. 

\subsection{Existing Fraud Detection Solutions}
Existing approaches to addressing cryptocurrency fraud fall into two distinct categories: industrial solutions which generally emphasize SaaS (Software as a Service) or PaaS (Platform as a Service) integrations and academic solutions which typically use open source tools and datasets. Each sector's predominant approach faces unique ongoing challenges. 

\subsubsection{Industrial Solutions}
Many current industrial solutions claim to investigate and detect cryptocurrency financial crime including TRM \cite{noauthor_trm_nodate}, feedzai \cite{noauthor_ai-powered_nodate}, and Chainalysis \cite{noauthor_blockchain_nodate}. While these smaller companies have grown their offerings, traditional sector leaders still play a central role. 
JPMorgan, for instance, has been a key player in recent efforts to combat cryptocurrency fraud. Although CEO Jamie Dimon has expressed a clear distaste for the medium that he once described as ``hyped-up fraud" \cite{sigalos_jamie_2024}, the company been a public proponent of blockchain technologies \cite{balevic_jamie_2025}. 

This support is most clear in their use of AI cryptocurrency fraud detection tools that use a combination of neural networks, deep learning, natural language processing, reinforcement learning, and computer vision to catch fraudulent behavior as well as detect both traditional and DeFi fraud. JPMorgan reports that the use of AI-powered mechanisms has saved them at least \$250 million annually and enabled the company to more effectively respond to customer reports \cite{tulsi_transforming_2024}. 

Despite the great success of the program, JPMorgan has suffered greatly from the black-box nature of their models. Due to data imbalances, anomaly detection models have a tendency to yield high false positive rates, resulting in the unintentional targeting of innocent customers. The company also faces continuous challenges integrating new systems into legacy platforms and regulatory uncertainty. Alternative tools, such as Chainalysis, also boast strong solutions yet continue to be hindered by challenges similar to those faced by major financial institutions \cite{narula_chainalysis_2024}. 

\subsubsection{Academic Solutions}
There are many ongoing academic research efforts that can guide the search for cryptocurrency fraud solutions, both directly and indirectly. 

Machine learning (ML) has grown to play a critical role in the identification and prevention of fraudulent activities. One such prominent application is credit card fraud detection \cite{bin_sulaiman_review_2022}. While traditional methods for credit card fraud detection have been in place for nearly as long as credit cards, the growing interest in ML has naturally led to its application to the field. As credit card fraud detection is similar in theme and nature to cryptocurrency fraud detection, the problems and solutions addressed by ML approaches can likely be (at least partially) transferred between the two.  

Beyond applying known application methods to new problems, there is a growing body of research surrounding cryptocurrency-specific approaches to fraud detection. Similar to the research surrounding credit card fraud detection, supervised and semi-supervised models combined with hybrid learning approaches tend to produce the most accurate fraud detection models \cite{bhowmik_comparative_2021}. Other researchers, however, have begun to propose novel ML models that more effectively address the graphical, interconnected nature of blockchain transaction data \cite{aziz_lgbm_2022}. 

While ML has been leveraged to perpetrate cryptocurrency scams, promising research indicates that these tools can also be part of a solution \cite{ashfaq_machine_2022}. Traditional data analytics tools, along with more advanced methods, rely heavily on the more straightforward nature of both structured and unstructured datasets. The study of graph data (characterized by a network of nodes and edges storing information in both the instance itself and the connection between instances) is still relatively new, as the advantage of this dataset style became most prominent after the rise of the internet. As blockchain data is best presented as a graphical dataset, traditional data analytics techniques have often fallen short as the nuances of transactions is lost. Machine learning models, however, excel in this domain, making them a promising tool for the job. Furthermore, developments in ML have suggested encouraging solutions to problems such as high false positive rates and real-time detection \cite{kathareios_catch_2017} through the addition of a nearest neighbor checker after the unsupervised anomaly detection stage. 

While many existing approaches to modernizing fraud detection focus on the models used to detect fraud alone, it should be noted that the explainability of these models is also necessary. There is a growing interest in combining XAI models and LLMs to improve the interoperability of black box models that could be the key to addressing concerns surrounding model faithfulness and explainability\cite{cambria_xai_2024}. Furthermore, this particular application of LLMs appears to be relatively successful in other applications and research \cite{wu_usable_2025}.

\subsection{Limitations of Existing Solutions}
To address the necessary features discussed in Section~\ref{sec: motivation}, a viable solution should:
\begin{enumerate}
    \item Use graphical data to fairly represent the nature of cryptocurrency transactions. 
    \item Integrate a graphical ML model for fraud detection (such as a GNN).
    \item Utilize an XAI explainer to address the black box nature of the fraud detection model.
    \item Integrate some kind of natural language explainer or other solution to allow non-technical users to understand the pipeline output.
    \item Conduct tests using data within the fraud domain.
    \item Classify the actual fraud type as different fraud types are denoted by unique behavior and have varying degrees of magnitude. 
\end{enumerate}

As indicated in Table~\ref{tab:related_work}, at the time of writing, there is very little research that combines a graphical database, GNN, XAI, and LLM for human-understood fraud detection. The most closely-related such work is by Nicholls et al. which does combine a GNN, XAI explainer, and LLM explanation to perform fraud detection \cite{nicholls_enhancing_2023}. Yet this research is primarily centered around building transaction \textit{narratives} rather than identifying types of fraud. The proposed work here expands beyond this approach in a few ways. First, by integrating a more refined LLM prompting strategy informed by existing research \cite{zytek_llms_2024}. Second, the modular and open source nature of the work allows for more flexible refinement of the pipeline to meet individual needs. Third, the addition of a user interface further centralizes the human accessibility of the model output. Finally, the proposed model is used to build a fraud classification rule-book which can be used to automate fraud classification in future research. This extends existing work meaningfully, as it allows future research to extend beyond binary fraud classification.

\begin{table*}[h!]
\centering
\caption{Comparison of related work in AI-based fraud detection. Our work is the first (to our knowledge) to combine graph-based databases, GNN anomaly detection, an XAI explainer, and an LLM-generated explanation layer.}
\label{tab:related_work}
\begin{tabular}{|p{6cm}|c|c|c|c|c|c|}
\hline
\textbf{Work} & \textbf{Graph Data} & \textbf{GNN} & \makecell{\textbf{XAI} \\ \textbf{Explainer}} & \makecell{\textbf{LLM} \\ \textbf{Explanations}} & \textbf{Fraud Domain}& \makecell{\textbf{Fraud Type} \\ \textbf{Classification}} \\
\hline
Kothapalli et al. (2024) \cite{kamisetty_deep_2021} \newline 
\textit{Deep Learning for Fraud Detection in Bitcoin Transactions} 
& \xmark & \xmark & \xmark & \xmark & \cmark & \xmark\\
\hline
Stephe et al. (2024) \cite{stephe_blockchain-based_2024} \newline 
\textit{Blockchain-Based Private AI Model with RPOA Based Sampling Method for Credit Card Fraud Detection} 
& \cmark & \xmark & \xmark & \xmark & \cmark & \xmark\\
\hline
Balusamy et al. (2025) \cite{balusamy_protecting_2025} \newline 
\textit{Protecting Financial Transactions and Cryptocurrency Networks from Fraud Using AI-Powered Blockchain Technology} 
& \cmark & \xmark & \xmark & \xmark & \cmark & \xmark\\
\hline
Balusamy et al. (2026) \cite{balusamy_protecting_2025} \newline 
\textit{Protecting Financial Transactions and Cryptocurrency Networks from Fraud Using AI-Powered Blockchain Technology} 
& \cmark & \xmark & \xmark & \xmark & \cmark & \xmark\\
\hline
Dhieb et al. (2020) \cite{dhieb_secure_2020} \newline 
\textit{Protecting Financial Transactions and Cryptocurrency Networks from Fraud Using AI-Powered Blockchain Technology} 
& \cmark & \xmark & \xmark & \xmark & \cmark & \xmark\\
\hline
Taher et al. (2024) \cite{taher_advanced_2024} \newline 
\textit{Advanced Fraud Detection in Blockchain Transactions: An Ensemble Learning and Explainable AI Approach} 
& \xmark & \xmark & \cmark & \xmark & \cmark & \xmark\\
\hline
Kapale et al. (2024) \cite{kapale_explainable_2024} \newline 
\textit{Explainable AI for Fraud Detection: Enhancing Transparency and Trust in Financial Decision-Making} 
& \xmark & \xmark & \cmark & \xmark & \cmark & \xmark\\
\hline
Li et al. (2025) \cite{li_hybrid-llm-gnn_2025} \newline 
\textit{Hybrid-LLM-GNN: integrating large language models and graph neural networks for enhanced materials property prediction} 
& \cmark & \cmark & \xmark & \cmark & \xmark & \xmark\\
\hline
Baghersashi et al. (2025) \cite{baghershahi_nodes_2025} \newline 
\textit{From Nodes to Narratives: Explaining Graph Neural Networks with LLMs and Graph Context} 
& \cmark & \cmark & \xmark & \cmark & \xmark & \xmark \\
\hline
Cedro et al. (2025) \cite{cedro_graphxain_2025} \newline 
\textit{GraphXAIN: Narratives to Explain Graph Neural Networks} 
& \cmark & \cmark & \cmark & \cmark & \xmark & \xmark\\
\hline
Nicholls et al. (2023) \cite{nicholls_enhancing_2023} \newline 
\textit{Enhancing Illicit Activity Detection using XAI: A Multimodal
Graph-LLM Framework} 
& \cmark & \cmark & \cmark & \cmark & \cmark & \xmark\\
\hline
\textbf{This Work} (2025)  \newline 
\textit{Explain First, Trust Later: LLM-Augmented
Explanations for Graph-Based Crypto Anomaly
Detection} 
& \cmark & \cmark & \cmark & \cmark & \cmark & \cmark\\
\hline
\end{tabular}
\end{table*}

\section{Proposed System Architecture}
\begin{figure*}
    \centering
    \includegraphics[width=1.0\linewidth]{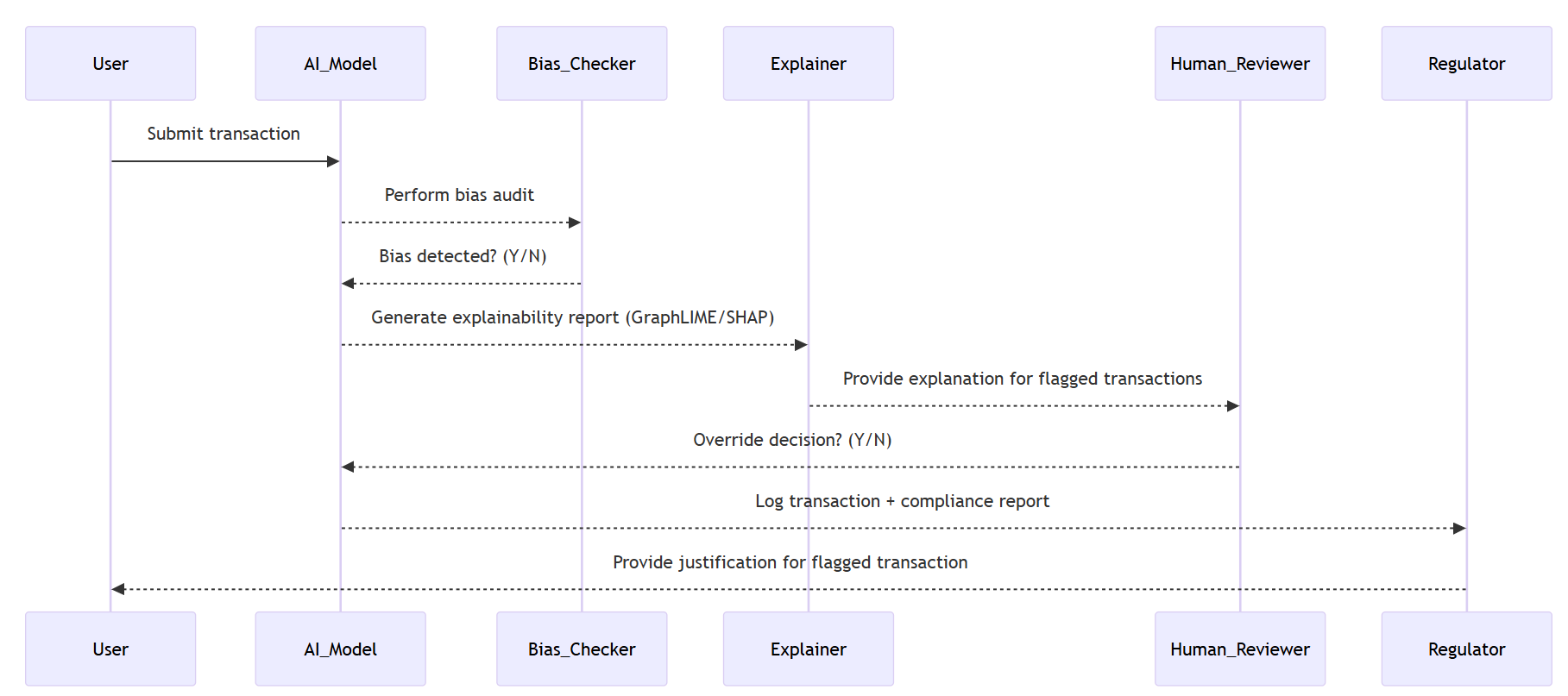} 
    \caption{A Sequence Diagram of the Proposed Solution}
    \label{fig:SeqDiagram}
\end{figure*}

As shown in Figure \ref{fig:SeqDiagram}, a solution that addresses many of the limitations faced by existing solutions requires a system of checks and balances at each stage of the cryptocurrency transaction and detection process. The proposed model is trained on graph-based transaction data to address limitations in existing research, enabling the model to detect fraud occurring from any coin. Additionally, an XAI model is run on the trained fraud detection model to explain the instances of fraud detected. This addresses the black-box nature of the model by exposing the features used to make a particular determination. A human-in-the-loop audit enhanced by LLM-generated explanations is performed on fraudulent instances as a final step before the case, with the explanations generated by the proposed process, can be taken to regulators or enforcers. Depending on the implementation, an enforcement party could use this information to, for example, automatically shut down accounts and produce a report for human reviewers if the action is appealed. 

The proposed multistage solution allows each instance of the fraud detection process to be verified and documented. Additionally, the solution is legislatively agnostic and modular; thus, it could be adapted to conform to regulations from any area or multiple areas at once. The solution could also be expanded to integrate a Retrieval-Augmented Generation (RAG) model between the XAI and human reviewer phase to cross-compare transaction behavior with existing legislative policy documents. 

\section{Implementation}
The research presented in this paper provides a working solution for all non-human components discussed above.\footnote{The implementation of the regulation integration components of the workflow require real-world implementation and testing  and are thus left for future research.} The proposed pipeline is designed to be model-agnostic: feature attributions and the LLM stage operate exclusively on (i) the model’s scalar node anomaly score and (ii) node feature values. This means the pipeline can be applied to any GNN or classical graph anomaly detector that returns a per-node score. Alternative XAI and LLM models can also be integrated with small changes to the prompt to provide context regarding the XAI semantic interpretation as needed. This solution was executed in four distinct modular stages: 
\begin{enumerate}
    \item Anomaly detection: A GNN (Graph Neural Network) is trained using unsupervised learning to detect anomalies in the graph dataset. 
    \item Explainer: GraphLIME is used to produce explanations for specified anomalies detected by the trained GNN model. 
    \item Human Readable Explanations: The XAI output and node features are passed to an LLM model to produce a human readable fraud explanations. 
    \item User Interface: Optionally the graph data is combined with the generated explanations and XAI scoring to generate a user interface that displays the information in a more human-readable format. 
\end{enumerate}

The final solution advances existing research by Hasan et al. by adding an LLM explanation layer and user interference to improve the output \cite{hasan_detecting_2024}. Furthermore, the addition of distinct modular phases and human-centric design distinguishes it from the framework proposed by Nicholls et al. \cite{nicholls_enhancing_2023} and similar binary classification solutions. Detailed documentation of each step follows. 

\subsection{Anomaly Detection}
To establish a graph-based cryptocurrency database as specified in the framework, the Elliptic++ Dataset, which is composed of 203,000 Bitcoin transactions and 822,000 wallet addresses, was used \cite{elmougy_demystifying_2023}. Nodes in the dataset, which represented wallets, contained information such as the number of transactions, transaction totals, and bitcoin sent and received. Edges, which represented transactions, contained information such as mean, median, and maximum bitcoin sent through the edge. A graph-structured transaction dataset of cryptocurrency transactions was produced by joining the node and edge (wallet and transaction) tables. The graph was then used to train an unsupervised Graphical Neural Network (GNN) to detect anomalies following Algorithm~\ref{alg:GNN}.

\begin{algorithm}[h!]
\caption{AnomalyGCN for Unsupervised Fraud Detection}
\KwIn{Graph $G = (V,E)$, node features $X$, epochs $T$, learning rate $\eta$, anomalies $k$}
\KwOut{Top-$k$ anomalous nodes with scores}

\BlankLine
\textbf{Model:} AnomalyGCN
\begin{itemize}[noitemsep,left=0pt]
  \item Layer 1: GCNConv($d_{\text{in}} \to h$), ReLU
  \item Layer 2: GCNConv($h \to 1$) \hfill (per-node anomaly score)
\end{itemize}

\BlankLine
\textbf{Training:}
\For{$t = 1 \dots T$}{
  Compute scores $s \gets f_\theta(X,E)$ \\
  Target = mean of $s$ \\
  $\mathcal{L} \gets \text{MSE}(s, \text{Target})$ \\
  Update $\theta \gets \theta - \eta \nabla_\theta \mathcal{L}$
}

\BlankLine
\textbf{Anomaly Detection:}
\begin{enumerate}[noitemsep,left=0pt]
  \item Compute final scores $s_i$ for all nodes $v_i \in V$
  \item Assign anomaly label: $y_i = \mathbb{1}[s_i < \mu_s - 2\sigma_s]$
  \item Rank nodes by $s_i$ and return top-$k$ anomalies
\end{enumerate}
\label{alg:GNN}
\end{algorithm}

Anomaly scoring is framed as an unsupervised deviation-from-typical-behavior objective. The GNN is trained to output a single node score and optimized with a mean-anchored MSE loss. Under this objective, nodes whose learned score is substantially below the population mean indicate that the model assigns them behavior different from the bulk of nodes and thus ranks them as anomalous. This objective is preferable in the presented use case as (1) it is model-agnostic and stable to implement across GNN backbones, (2) it avoids constructing explicit negative examples which are typically unavailable in fraud datasets, and (3) it yields an interpretable continuous anomaly ranking system to downstream XAI and LLM narration.

The model was configured as a two-layer Graph Convolutional Network (GCN) using PyTorch Geometric using the following parameters:  
\begin{itemize}
    \item \textbf{Architecture:} GCNConv–ReLU–GCNConv with a 16-dimensional hidden layer, outputting one anomaly score per node.
    \item \textbf{Input Features:} 25 blockchain attributes including graph structure (degree, in/out-degree), transaction behavior (sent/received counts and volumes), temporal activity (lifetime, block intervals), and address diversity.
    \item \textbf{Training Objective:} Self-supervised regression toward the global mean score using MSE; anomalous nodes deviate significantly from the learned pattern.
    \item \textbf{Optimization:} Adam (lr=0.01), 350 epochs.
    \item \textbf{Labeling Rule:} Nodes flagged as anomalous if their score lies below $\mu - 2\sigma$.
\end{itemize}

 A GNN was selected as comparable literature has revealed strong results for anomaly detection applications similar to the one presented in this paper \cite{li_hybrid-llm-gnn_2025} \cite{cedro_graphxain_2025} \cite{nicholls_enhancing_2023}. \footnote{While existing literature indicates that ensemble methods may present stronger results for anomaly detection \cite{oliveira_guiltywalker_2021}\cite{weber_anti-money_2019}, the use of a GNN in this context is most appropriate as it more directly addresses the relational nature of cryptocurrency transaction data. Fraud signals often manifest in local subgraph structures \cite{zhang_hidden_2017}, which GNNs are well-suited to capture through neighborhood aggregation \cite{xie_when_2020}. Moreover, GNN embeddings support unsupervised anomaly detection \cite{wang_decoupling_2021} and integrate directly into XAI models, which is not possible for ensemble approaches.} The result of this process was an updated graph database that included an anomaly score as well as the trained GNN model. As the baseline dataset contained insufficient labels for the desired outcome, the model was trained using unsupervised learning. The model output was, however, later compared to the machine learning generated labels provided in the Elliptic++ dataset  \cite{elmougy_demystifying_2023}. 

\subsection{XAI Explainer}
The GNN model was then interpreted using GraphLIME, which fits a local linear surrogate model to estimate feature-level importances for each node. GraphLIME was selected because the proposed use case, prompting an LLM to reason about transaction behavior, requires human-interpretable feature attributions, which aligns with the design of GraphLIME \cite{huang_graphlime_2023}.\footnote{This choice does not conflict with the model-agnostic nature of the pipeline but rather supports it. GraphLIME, like LIME and KernelSHAP, is a post-hoc, model-agnostic explainer: it requires only node features and a scalar anomaly score at each node. It applies equally well to any anomaly model (GNN, autoencoder, embedding-based model, or rule-based scoring function).} Although GraphLIME provides the cleanest interface for the LLM stage, the pipeline can integrate alternative explainers when appropriate by modifying the LLM prompt to clarify the semantics of their importance values. 

To produce the best output, GraphLIME was configured as follows: 
\begin{itemize}
    \item \textbf{Subgraph Extraction:} $2$-hop ego networks (capped at 2000 nodes) using PyG's \texttt{k\_hop\_subgraph}.
    \item \textbf{Standardization:} Subgraph features were z-normalized prior to explanation.
    \item \textbf{Model:} Positive Lasso regression over a Gaussian kernel similarity matrix was used to obtain local linear feature weights. This reveals which features contributed the most to a node being highlighted as an anomaly.  
    \item \textbf{Selection:}  Each feature is given a feature weight that was then used to rank the most influential features. This information naturally implies possible fraudulent activity.  The top 3 non-zero features (importance $>10^{-8}$) were retained.
    \item \textbf{Output:} A structured explanation combining feature weights and wallet statistics for downstream LLM reasoning was returned. 
\end{itemize}

\subsection{LLM Explanations}

After the GraphLIME model had been trained, this output combined with the original node features was used as the input prompt for the LLM model to generate human-readable explanations. The prompt, shown below, included context for the task, variables where the node features and GraphLIME weights would be added, and the output structure syntax. This follows best practices for combined XAI and LLM tasks based on the limited research available \cite{li_hybrid-llm-gnn_2025}. The LLM parameters are as follows:
\begin{itemize}
    \item \textbf{Model:} \texttt{gpt-4o-mini}, temperature $=0.2$ for reproducibility.
    \item \textbf{Output Format:} Strict JSON containing:
    \begin{verbatim}
{
  "explanation": "... (2-5 sentences)",
  "is_fraud": true/false,
  "fraud_type": "... or null",
  "confidence": float,
  "evidence": {{
      "features": list of features,
      "behaviors": list of patterns
}
    \end{verbatim}
    \item \textbf{Fraud Taxonomy:} Restricted to a fixed list, shown below.
    \item \textbf{Robustness:} Exponential-backoff retry mechanism (up to 12 attempts) to mitigate rate limits and network errors.
    \item \textbf{Metrics Logged:} LLM latency, prompt/completion token counts, and estimated cost (USD).
\end{itemize}

The full LLM API call is as follows: 

\begin{lstlisting}[caption=LLM Input Script]
You are a cryptocurrency forensics analyst.

You will receive:
1. GraphLIME feature importances
2. Raw node statistics
3. Fraud-type ontology

Return STRICT JSON:

{{
  "explanation": "... (2-5 sentences)",
  "is_fraud": true/false,
  "fraud_type": "... or null",
  "confidence": float,
  "evidence": {{
      "features": [list of most relevant feature names],
      "behaviors": [list of notable behavioral patterns]
  }}
}}

Rules:
- fraud_type MUST be from: {fraud_types}
- If not fraudulent → fraud_type = null.
- Be concise and data-grounded.
- Return ONLY JSON.

### Feature Importances
{formatted_weights}

### Node Statistics
{formatted_data}
\end{lstlisting}

The input for the prompt included the top three features identified by GraphLIME as well as the primary node features. A sample of the information automatically added to the prompt is shown below.

\begin{lstlisting}[caption=Sample Input]
Node ID: "36Wu3jDLUgD2suGdEsZkHycu9htUm8JfCg":
Top Features:
    - "degree": 4
    - "num_txs_as_sender": 2.0
    - "num_txs_as_receiver": 0.0
    - "total_txs": 2.0
    - "btc_sent_total": 0.94528131
    - "btc_received_total": 0.0,
GraphLIME Scores: 
    - "btc_received_mean": 0.9937777519226074
    - "degree": 0.9937777519226074
    - "blocks_btwn_input_txs_mean": 0.24844442307949066
    - "num_addr_transacted_multiple": 1.4437569362257818e-08
    - "transacted_w_address_mean": 1.7219119463618426e-09
\end{lstlisting}

The LLM input was also provided with a list of possible fraud types based on those identified by Trozze et al. \cite{trozze_cryptocurrencies_2022} including:
\begin{itemize}
    \item Ponzi schemes
    \item Phishing attacks
    \item Pump-and-dump schemes
    \item Ransomware
    \item SIM swapping
    \item Mining malware
    \item Giveaway scams
    \item Impersonation scams
    \item Securities fraud
    \item Money laundering
\end{itemize}
Notably, not all of these schemes will be detectable from the transaction data as fraud attacks like SIM swapping, malware mining, and impersonation scams likely appear normal from the transaction data alone. 

Based on the prompt and information provided, the LLM generated an evaluation regarding the likelihood of fraud, fraud type, a brief explanation for the conclusion, and a confidence score. A sample of a single LLM output from the node presented earlier is provided below. 

\begin{lstlisting}[caption=Sample Output]
"explanation": "The analysis indicates that the wallet has a high degree of transactions with a mean of 0.945 BTC sent, but no BTC received. The significant anomaly score suggests unusual behavior, particularly with the high blocks between transactions. This pattern is indicative of potential money laundering activities.",
"is_fraud": true,
"fraud_type": "money laundering",
"confidence": 0.85,
"evidence": {
  "features": [
    "btc_sent_mean",
    "degree",
    "blocks_btwn_input_txs_mean"
  ],
  "behaviors": [
    "High transaction volume with no incoming funds",
    "Long intervals between transactions"
\end{lstlisting}

The LLM was polled 5 times for each node and the scores were averaged to mitigate LLM hallucination and to support reproducibility and repeatability. These results were merged to create a consensus which took the majority label for the binary "is\_fraud" label and the fraud type. The agreement ratio for the binary and type label as well as the average confidence were also calculated. A sample of the LLM consensus output is shown below. 

\begin{lstlisting}[caption= Node Sample Consensus Sample] 
"consensus": {
    "node_id": "36Wu3jDLUgD2suGdEsZkHycu9htUm8JfCg",
    "is_fraud": true,
    "fraud_type": "money laundering",
    "agreement_rate": 0.6,
    "fraud_type_agreement": 1.0,
    "avg_confidence": 0.51
  },
\end{lstlisting} \label{lst:llm-consensus}

In this example, one can observe that the majority of the models (3 out of 5 as the agreement rate is 0.6) agreed that the node was fraudulent. Of the 3 models that determined the nodes was fraudulent, all 3 agreed that the fraud type was money laundering (indicated by the fraud type agreement of 1). The average confidence of this determination was 0.51, a relatively low score which is logical since not all models agreed that the node was fraudulent. 

\subsection{Interactive UI}
Finally, as proof-of-concept, a small subset of the database combined with associated generated explanations was used to create an interactive User Interface (UI) to visually explore the interactions between wallets (nodes) and transactions (edges). As the dataset is fairly large, it would not be feasible for the full dataset to be displayed. Examples of the UI are shown in \ref{fig:DashExample1} as well as in appendix \ref{sec:DashExamples}.

\begin{figure}[h]
    \centering
    \includegraphics[width=1.0\linewidth]{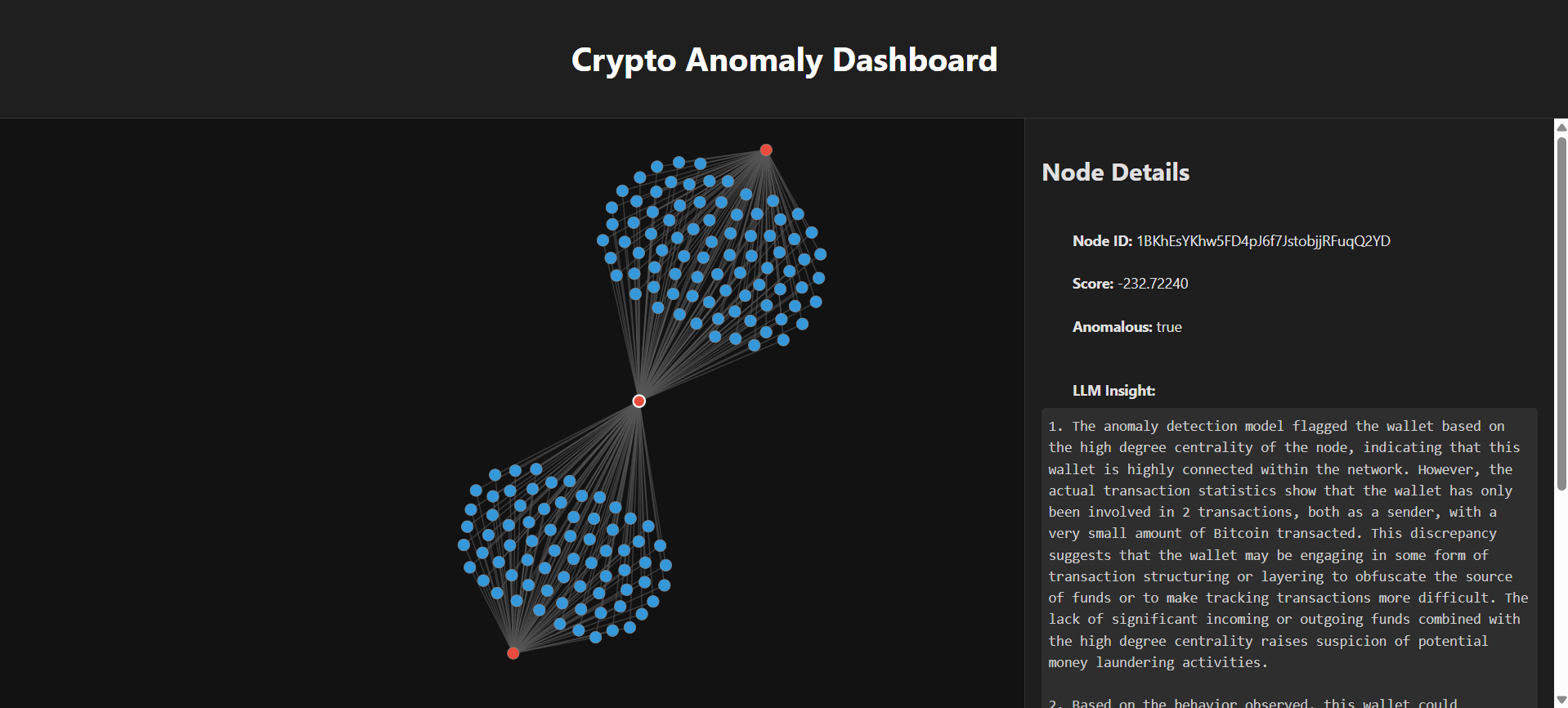} 
    \caption{Sample Dashboard with LLM Explanations, Restricted to a Single Fraudulent Wallet}
    \label{fig:DashExample1}
\end{figure}

As shown, the UI allows users to select nodes of interest. Fraudulent nodes are displayed in red while benign nodes are blue. As not all nodes have produced explanations, since this would be too costly to compute for the entire dataset, nodes with an existing LLM explanation are outlined in white. When a user selects a node, the Node details that were provided to the LLM along with the anomalous score generated at the GCN stage and the LLM insight (if available) are displayed in the panel on the right. 

\section{Discussion of Results}
The final output of the system clearly identifies anomalies and provides appropriate explanations for the reasoning behind the label. While scaling the full pipeline presents a clear challenge, as the size of graph based data in particular tends to be exponential in nature, the presented pipeline can be implemented in a variety of settings to add explainability to black box ML solutions. 

This section presents a qualitative and quantitative evaluated individually to ensure the faithfulness and usability of the pipeline. The LLM and XAI output was assessed using the full dataset. Then, a sample of the top 250 anomalies with their associated XAI weights and LLM outputs were used to evaluate the LLM outputs and full pipeline. 

\subsection{Iterative Prompting Results}

To further assess the contribution of the LLM layer, the iterative improvements made through the prompt engineering process were documented. Each modification to the prompt was designed to increase interpretability, moving from raw GraphLIME feature weights to explanations that align with standard forensic investigation practices (cite?). Table~\ref{tab:prompt_improvements} summarizes these iterations, showing how progressively richer prompts led to clearer and more grounded natural language outputs.

\begin{table*}hb!] 
\centering
\caption{Impact of prompt modifications on LLM-generated explanations. Each iteration progressively improved interpretability and alignment with forensic reasoning.}
\label{tab:prompt_improvements}
\begin{tabular}{p{3cm}p{4cm}p{6cm}p{3cm}}
\hline
\textbf{Prompt Modification} & \textbf{Example Node Features} & \textbf{LLM Output (excerpt)} & \textbf{Interpretability Gain} \\
\hline
Baseline & degree: 0.99, btc\_sent\_total: 0 & ``Suspicious due to degree centrality'' & Minimal context; vague reasoning \\

+ Node statistics & + total\_txs, btc\_received\_total & ``High inflows without proportional outflows; possible laundering'' & Clearer forensic pattern identified \\

+ Fraud types in prompt & + fraud type list & ``Behavior resembles Ponzi scheme (many inputs, no outputs)'' & Explicit link to known fraud typology \\

+ Clarify weights vs. values & Explicitly separate GraphLIME weights vs. true values & ``Although GraphLIME weighted `btc\_sent\_total`, the true value is 0, suggesting hoarding'' & Improved alignment between features and narrative \\

+ Few-shot examples & Provided sample labeled cases & ``High degree + quick fund turnover → consistent with layering stage of money laundering'' & Most nuanced, regulator-ready explanation \\
\hline
\end{tabular}
\end{table*}

As a secondary contribution of this work, the iterative prompt engineering process revealed several key findings regarding improving LLM outputs for GNN + XAI systems: 
\begin{itemize}
    \item Ensuring that the LLM can distinguish between the GNN node features and the XAI feature weights can be difficult. The two components must be explicitly stated and differentiated to produce useful outputs.
    \item Adding categorization details (when possible), such as types of fraud, provides a better baseline for the LLM to translate between numeric values and real world implications. 
    \item Few-shot prompting does improve the LLM output in terms of quality, succinctness, and reproducibility. However, this strategy also significantly increased the input token count. 
\end{itemize}

\subsection{Qualitative Validation}
\subsubsection{Anomaly Detection Model}
The anomaly detection model was evaluated using the labels provided in the Elliptic++ dataset. While these cannot be considered ground truth labels, as they are also ML generated, the analysis does provide a baseline comparison for similar models on the same dataset. To convert the anomaly score into a binary decision, a 98\% threshold cutoff was used where any nodes with an anomaly score in the top 98th percentile were marked as anomolies. The labels produced by the GNN trained in this paper and the labels provided by the authors of the Ellipticc++ dataset agree on 75\% of labels. Interestingly, the GNN and LLM decisions agree on only 30\% of labels given the same threshold. An example of this failure mode, where the GNN and LLM disagree, is provided in Section~\ref{sec: full-llm-examples}. As the threshold is lowered, the dataset and GNN labels disagree more while the LLM and GNN labels agree more. 

\subsubsection{XAI Model Faithfulness}
The GraphLIME faithfulness was analyzed using a global surrogate model. A ridge regression model was trained on the dataset to predict the node’s anomaly score from the node features. The features were standardized so that coefficients were comparable. A large absolute coefficient implies that feature strongly influences anomaly score. Both methods indicated that the GNN focuses on transaction-related features, especially BTC sent/received totals and means. Temporal features like block intervals were also highly influential. The two methods largely agreed, with a spearman correlation coefficient of 0.497 and a p-value of p=0.0135, validating the XAI model’s reasoning.

\subsubsection{LLM Validation and Output Observations} \label{sec: llm-output}
\begin{figure}[h!]
    \centering
    \includegraphics[width=1.0\linewidth]{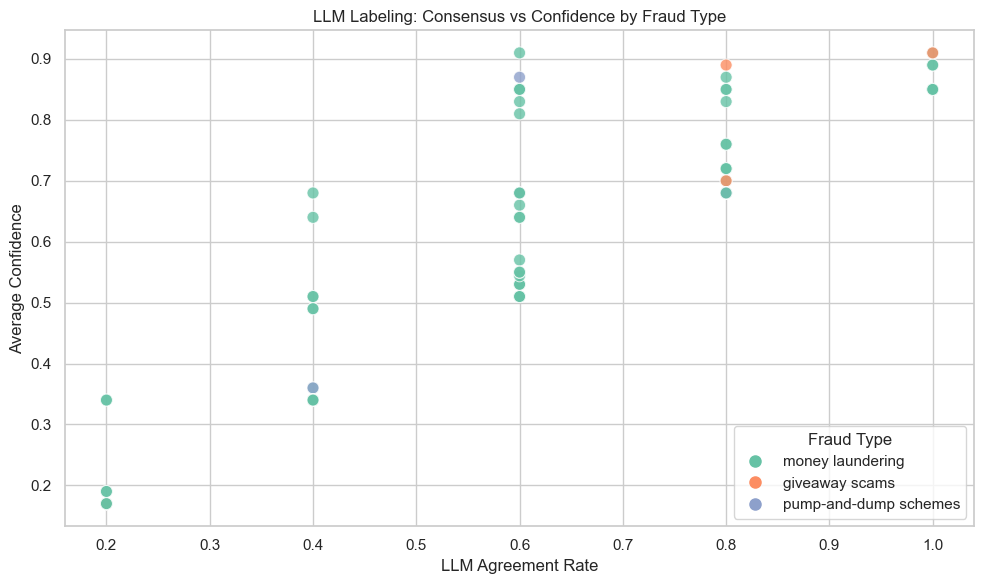} 
    \caption{Comparison of LLM consensus on the same node, confidence, and fraud type classification.}
    \label{fig:LLM-Consensus}
\end{figure}

To minimize the risk of miscategorization due to LLM hallucination, each LLM analysis was repeated 5 times per node with a temperature of 0.2. The LLM consensus scorecard shown in \ref{lst:llm-consensus} was then constructed using the majority vote of the LLM determinations and their agreement. Figure~\ref{fig:LLM-Consensus} demonstrates the overall trend of LLM consensus. 

As expected, as the LLM confidence in the classification increases, the consensus also increases. Although giveaway scams and pump-and-dump schemes were less present in the data, They both yielded high agreement and confidence when they were selected. 

\subsection{Model Cost and Latency}
Next, we share results from a cost and latency comparison in \ref{tab:cost-comparison}, which demonstrates the impact of each stage on the overall pipeline runtime and price. Notably, as expected, the explainer dominates the runtime, highlighting a known disadvantage of using "heavy explainer" such as GraphLIME in larger systems. The LLM input token usage is high, which could limit applicability, but newer research in LLM prompt caching could improve this metric \cite{potamitis_cache_2025}.
\begin{table}[H]
\centering
\caption{Summary Statistics for GNN, Explainer, and LLM Performance}
\label{tab:cost-comparison}
\begin{tabular}{lrrrr}
\hline
\textbf{Metric} & \textbf{Count} & \textbf{Mean} & \textbf{Median} & \textbf{Sum} \\
\hline
GNN time (s)              & 201 & 920.679  & 917.470  & 1.85$\times 10^{5}$ \\
Explainer time (s)        & 201 & 2642.829 & 2605.739 & 5.31$\times 10^{5}$ \\
LLM latency (s)           & 201 & 4.086    & 3.588    & 8.21$\times 10^{2}$ \\
LLM cost (USD)            & 201 & 0.001247 & 0.001245 & 2.51$\times 10^{-1}$ \\
LLM input tokens          & 201 & 5411.592 & 5420.000 & 1.09$\times 10^{6}$ \\
LLM output tokens         & 201 & 725.080  & 721.000  & 1.46$\times 10^{5}$ \\
Total pipeline time (s)   & 201 & 2663.281 & 2629.415 & 5.35$\times 10^{5}$ \\
\hline
\end{tabular}
\end{table}

\subsection{Fraud Type Classification}

\begin{figure*}[b!]
    \centering
    \includegraphics[width=1.0\linewidth]{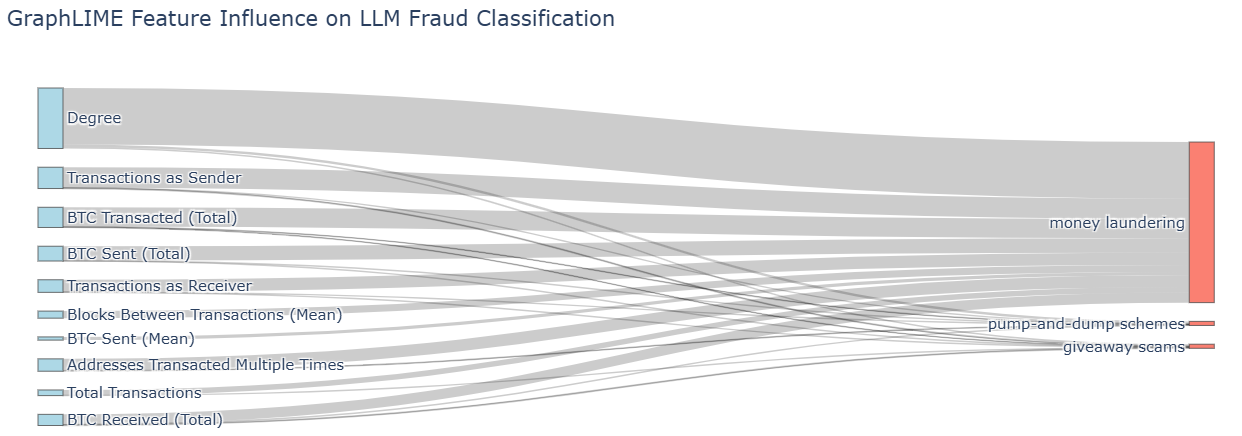} 
    \caption{A Sankey diagram demonstrating the connection between top GraphLime features and the fraud type assigned by the LLM.}
    \label{fig:SanKey}
\end{figure*}

In addition to the tabular comparison, a visual demonstration of interpretability gains can be seen by contrasting GraphLIME feature outputs with the corresponding LLM-generated explanations. This demonstrates how the XAI model contributed towards the LLM's decision making process. Figure~\ref{fig:SanKey} shows a Sankey diagram depicting the frequency and connection between the features marked as highly important and the LLM-based fraud classification. On the left, key features with high GraphLIME scores (e.g., degree centrality, transaction volume) are shown, while the right side presents fraud categories (e.g., Ponzi schemes, phishing, money laundering) selected by the LLM. The flow thickness corresponds to how frequently a feature was associated with a given fraud type across all flagged nodes. The graph clearly shows that the most frequently flagged fraud type was money laundering. Degree was the most commonly flagged reason across all classifications although some transaction features, such as the transaction total, mean, and received were also significant in several classifications. 

\begin{figure*}[h!]
    \centering
    \includegraphics[width=1.0\linewidth]{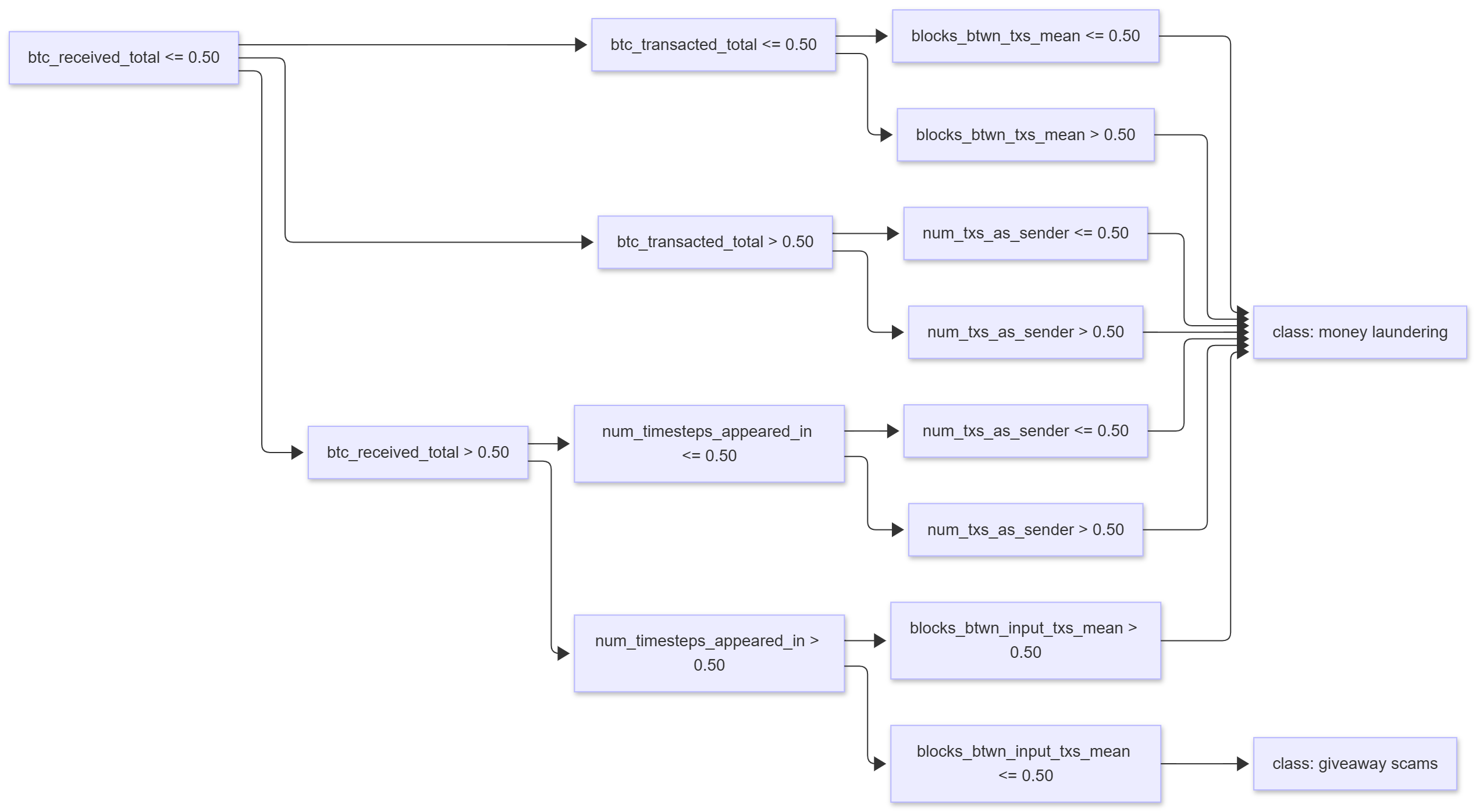} 
    \caption{Decision tree trained on the LLM-produced data to classify fraud types based on cited features.}
    \label{fig:DecisionTree}
\end{figure*}

To demonstrate how the analysis produced by the pipeline can be applied, the relationship between the predicted fraud type and model features was then used to build a decision tree. The decision tree was trained on the LLM classifications and reasons for classifications with a maximum depth of 3 to support interoperability. The trained model was tested on a small test set of the data and returned an accuracy of 0.8. The creation of the decision tree, shown in Figure~\ref{fig:DecisionTree} expands the scope of existing work by creating a pathway for multivariate labeling rather than binary classification \cite{nicholls_enhancing_2023}. Furthermore, it demonstrates the value of the LLM integration, as hand labeling the data would require extensive expertise in the field and may be fairly subjective. The integration of an LLM stage to provide labels thus automates this process reducing the human workload to verifying the natural language reasoning rather than interpreting raw data.  As expected, since money laundering schemes dominated the dataset, this is the most common selection, followed by giveaway scams. Although impersonation schemes and pump and dump schemes were identified in the sample dataset, they were not frequent enough to establish their place in the decision tree. It is worth noting that the decision tree was trained on a sample of only 250 nodes. On a larger scale, the decision tree would likely be significantly more detailed. 

\subsection{Transparency and Impact of LLM Usage}
The use of LLMs are integral to the proposed methodology, but their role is restricted to post-hoc explanation rather than anomaly detection itself. All prompts, few-shot examples, and generated text used in this study were documented and independently validated by the author. To ensure reproducibility, the input prompts and model outputs are provided in the paper. Furthermore, the code base, which includes the LLM setup and prompts, is publicly available and linked in Appendix~\ref{sec:CodeAvailabilty}. While a proprietary LLM (gpt 4.0-mini) was used, thus limiting reproducibility to an extent, the pipeline itself is model-agnostic and could be replicated with open-source alternatives such as LLaMA or Mistral. This explicit modularity mitigates concerns of vendor lock-in.

The use of LLMs naturally raises broader questions of responsible AI development. First, no additional data collection was conducted; the system operates entirely on publicly available blockchain transaction data (Elliptic++). Therefore, consent and data holder rights were not infringed nor would the pipeline ever require the use of private data as blockchain data is always publicly available and encrypted. Second, the need for an LLM was justified by the gap between technical feature weights and regulator-ready explanations as well as the difficult nature of the classification task. Without this translation layer, anomaly detection outputs remain opaque and unusable for forensic or legal purposes. Third, the environmental footprint of the experiments was minimized by (i) using the minimum viable number of API queries ($\sim$2k requests and $\sim$2.2M tokens across the study), (ii) relying on pre-trained models rather than fine-tuning, and (iii) executing all GNN and XAI training locally on a single GPU. This aligns with best practices for responsible ML experimentation as recommended by Lacoste et al.~\cite{lacoste_quantifying_2019}. 

\section{Limitations}
Most notably, the ability to validate the output of the pipeline is the largest limitation. Similar to other fraud detection and cyber attack research efforts, datasets that include ground truth are difficult to construct and thus challenging to find. Particularly in the application presented, a ground truth would necessitate positive confirmation of the fraudulent or non-fraudulent activity for every transaction to be truly viable. This limitation does, however bring out a strength in the paper as the output, while not verifiable, is comprehensible: human intuition can thus be used on a case by case basis to explore, refine, validate, or challenge AI-based predictions. Additionally, the introduction of a fraud classification scheme presented in this work contributes to ongoing efforts to combat these challenges. 

Further, because fraudulent behavior is, by nature, an anomaly in the dataset, model imbalances also raise concerns regarding the GCN model accuracy. While some of these concerns are addressed by the authors of the dataset and in the discussion section of this work, the true impact of this imbalance is difficult to measure. As discussed in Section~\ref{sec: llm-output} this problem, to an extent, may also be addressed by the addition of an LLM as a layer of context-aware verification is added. Furthermore, the integration of the XAI model alongside the LLM polling and justification provides the needed context behind model decisions. 

As the XAI explainer and LLM must generate explanations at the individual node level, the pipeline has a fairly high computational cost, particularly at larger scales. This also limits the usability of the pipeline as there is no way to batch the XAI and LLM processes with the current pipeline. 

Furthermore, the LLM prompt simply provides a list of cryptocurrency fraud types to the LLM rather than explicitly defining each type. The pipeline assumes that the LLM can distinguish between each fraud type, identify key behaviors of the fraud type, and connect them to the provided model outputs. As mentioned earlier, the resulting classifications and trends were used to create a more standard method of classifying fraud types based on node behavior. More research is needed, however, to further develop this topic as very little existing research addresses the deterministic classification of cryptocurrency fraud, including the prevalence of different fraud types and emergence of novel forms of fraud and patterns associated with them. 

Finally, as with many LLM centered pipelines, the balance between reproducibility and model originality is an ongoing challenge. To address reproducibility concerns, the model temperature was set relatively low (0.2) and the model was prompted multiple times for each node. 

\section{Future Work}
There are a variety of research areas that would further develop the work presented:
\begin{itemize}
    \item Generating LLM insights for non-anomalous nodes could provide further clarity into regular versus irregular behavior. 
    \item A human evaluation and human labeled validation would also add critical information regarding the quality and accuracy of the LLM output and the pipeline as a whole. This presents practical challenges as both labeling and evaluation require field experts and accepted fraud type definitions. Thus, the presented work takes one step towards this process by providing a fraud type labeling mechanism. 
    \item In addition to human evaluations, an ablation study surrounding the ease of human labeling and classification given different amounts of data (features only, features with XAI explanations, features with XAI and LLM explanations) would also provide valuable context for the direction similar work should pursue. 
    \item Adding data from other blockchain transactions (the dataset used only included Bitcoin transactions) would add a layer of complexity and improve the range of applications.
    \item Connecting the insights with a more RAG-like system to more carefully define fraud types would improve the LLM insights.
    \item Conducting qualitative research into enforcement buy-in for ML enhanced pipelines would provide relevance to the work as well as direction in applying the work. Furthermore, real-world testing generally would help to evaluate the entire proposed pipeline. 
    \item The fraud type classification analysis produced an interesting depiction of the relationship between database features and distinct fraud types. A larger sample set could reveal a richer decision tree that could be used to classify fraud types with  a deterministic, self explanatory process.
\end{itemize}

\section{Conclusion}
This work introduced a modular end-to-end pipeline that unifies graph neural networks, explainable AI methods, and large language models to generate interpretable fraud insights within cryptocurrency transaction networks. By combining GNN anomaly scoring, GraphLIME feature attribution, and LLM-based explanation polling, the system produces structured, human-readable narratives and supports the construction of a derivative, data-driven decision-tree classifier for fraud type characterization. The results are promising both qualitatively—through coherent, domain-aligned explanations—and quantitatively, with stable consensus patterns emerging across independent LLM samples.

The pipeline is presented as a proof of concept rather than a turn-key production system. Real-world deployment would require evaluation at significantly larger scales, consideration of adversarial robustness, and collaboration with regulatory and industry stakeholders. Nonetheless, this work fills an important gap in the literature by extending anomaly detection beyond binary fraud classification and demonstrating a viable, regulation-agnostic path toward automated fraud-type categorization. As cryptocurrency-related crime continues to grow in scale and sophistication, interpretable and extensible tools of this kind offer a critical first step toward practical and trustworthy forensic systems.

\bibliography{references.bib}

\section{Appendix}
\subsection{GPT Usage}
A GPT was used to generate mermaid graphs based on the following prompts (with minor modifications for formatting consistency): 

\begin{lstlisting}
Generate a Mermaid `sequenceDiagram` that visualizes the interaction between different stakeholders in an AI-based transaction monitoring system. The diagram should include the following participants and interactions:  

Participants:
- `User`: The entity submitting a transaction  
- `AI_Model`: The automated system analyzing the transaction  
- `Bias_Checker`: A module that audits for bias  
- `Explainer`: A system generating explainability reports (GraphLIME/SHAP)  
- `Human_Reviewer`: A reviewer who can override AI decisions  
- `Regulator`: A compliance entity receiving logs and providing justifications  

Interactions:
1. `User` > (`Submit transaction`) > `AI_Model`  
2. `AI_Model` > (`Perform bias audit`) > `Bias_Checker`  
3. `Bias_Checker` > (`Bias detected? (Y/N)`) > `AI_Model`  
4. `AI_Model` > (`Generate explainability report (GraphLIME/SHAP)`) > `Explainer`  
5. `Explainer` > (`Provide explanation for flagged transactions`) > `Human_Reviewer`  
6. `Human_Reviewer` > (`Override decision? (Y/N)`) > `AI_Model`  
7. `AI_Model` > (`Log transaction + compliance report`) > `Regulator`  
8. `Regulator` > (`Provide justification for flagged transaction`) > `User`  

Ensure the output follows correct Mermaid syntax with appropriate message direction (`->>`, `-->>`) and clear labeling. The sequence should accurately reflect the decision-making and compliance process in transaction monitoring.  
\end{lstlisting}
This conversation can be found here: \url{https://chatgpt.com/share/67c8c692-aa9c-8001-9d74-35598d9c3a5f}

\subsection{Full Node Output Example} \label{sec: full-llm-examples}
Full node samples, with all 5 evaluations and the consensus is provided below (a portion of repeated data from the raw output has been removed to improve readability): 
\subsubsection{GCN and LLM Agree on Fraud Status}
\begin{lstlisting}
  "17dKfNjkUT9HGUS5hLStxMQyHD3NCF6vRY": {
    "top_features": [
      [
        "btc_sent_median",
        0.295053631067276
      ],
      [
        "btc_received_mean",
        0.13998724520206451
      ],
      [
        "blocks_btwn_output_txs_mean",
        0.0
      ],
      [
        "num_addr_transacted_multiple",
        0.0
      ],
      [
        "blocks_btwn_txs_mean",
        0.0
      ]
    ],
    "explainer_scores": {
      "btc_sent_median": {
        "graphlime": 0.295053631067276
      },
      "btc_received_mean": {
        "graphlime": 0.13998724520206451
      },
      "blocks_btwn_output_txs_mean": {
        "graphlime": 0.0
      },
      "num_addr_transacted_multiple": {
        "graphlime": 0.0
      },
      "blocks_btwn_txs_mean": {
        "graphlime": 0.0
      }
    },
    "llm_output_structured": {
      "samples": [
        {
          "explanation": "The analysis indicates a significant anomaly score of 78.98, suggesting unusual behavior for the wallet in question. The wallet has only transacted twice, with a very low total amount of BTC transacted, which is atypical for legitimate wallets. The lack of incoming transactions and the low frequency of activity raise suspicions of potential fraudulent behavior.",
          "is_fraud": true,
          "fraud_type": "money laundering",
          "confidence": 0.85,
          "evidence": {
            "features": [
              "btc_sent_median",
              "btc_received_mean",
              "anomaly_score"
            ],
            "behaviors": [
              "low transaction frequency",
              "minimal BTC transacted",
              "no incoming transactions"
            ]
          }
        },
        {
          "explanation": "The analysis indicates a high anomaly score, suggesting unusual behavior for the wallet in question. The wallet has only two transactions, with a significant gap between the first and last activity, which is characteristic of potential money laundering activities. The low volume of received transactions further supports this suspicion.",
          "is_fraud": true,
          "fraud_type": "money laundering",
          "confidence": 0.85,
          "evidence": {
            "features": [
              "btc_sent_median",
              "btc_received_mean",
              "anomaly_score"
            ],
            "behaviors": [
              "low transaction volume",
              "high anomaly score",
              "long inactivity period"
            ]
          }
        },
        {
          "explanation": "The analysis indicates a significant anomaly score of 78.98, suggesting unusual behavior for this wallet. The wallet has only two transactions, with a high median amount sent relative to the low amount received, which is atypical for legitimate activity. This pattern, combined with the lack of multiple address transactions and a high degree of isolation, raises concerns about potential fraudulent activity.",
          "is_fraud": true,
          "fraud_type": "money laundering",
          "confidence": 0.85,
          "evidence": {
            "features": [
              "btc_sent_median",
              "btc_received_mean",
              "anomaly_score"
            ],
            "behaviors": [
              "high median amount sent with low received",
              "limited transaction history",
              "high anomaly score indicating unusual activity"
            ]
          }
        },
        {
          "explanation": "The analysis indicates a significant anomaly score of 78.98, suggesting unusual behavior for this wallet. The wallet has only transacted twice, with a low total amount of BTC and no incoming transactions, which is atypical for legitimate wallets. These characteristics, combined with the lack of multiple address transactions, raise concerns about potential fraudulent activity.",
          "is_fraud": true,
          "fraud_type": "money laundering",
          "confidence": 0.85,
          "evidence": {
            "features": [
              "btc_sent_median",
              "btc_received_mean",
              "anomaly_score"
            ],
            "behaviors": [
              "low transaction volume",
              "no incoming transactions",
              "high anomaly score"
            ]
          }
        },
        {
          "explanation": "The analysis indicates a high anomaly score, suggesting unusual behavior for the wallet in question. The low number of transactions and the significant median of BTC sent imply potential manipulation or fraudulent activity. The wallet has only transacted twice, which is atypical for legitimate usage patterns.",
          "is_fraud": true,
          "fraud_type": "pump-and-dump schemes",
          "confidence": 0.85,
          "evidence": {
            "features": [
              "btc_sent_median",
              "btc_received_mean",
              "anomaly_score"
            ],
            "behaviors": [
              "low transaction frequency",
              "high anomaly score",
              "significant median BTC sent"
            ]
          }
        }
      ],
      "consensus": {
        "node_id": "17dKfNjkUT9HGUS5hLStxMQyHD3NCF6vRY",
        "is_fraud": true,
        "fraud_type": "money laundering",
        "agreement_rate": 1.0,
        "fraud_type_agreement": 0.8,
        "avg_confidence": 0.85
      },
      "agreement_rate": 1.0,
      "fraud_type_agreement": 0.8,
      "avg_confidence": 0.85,
      "avg_latency": 4.663082838058472,
      "total_cost_usd": 0.0012954,
      "total_input_tokens": 5420,
      "total_output_tokens": 804,
    "node_data": {
      "degree": 6,
      "in_degree": 0,
      "out_degree": 0,
      "node_type": "unknown",
      "type": "wallet",
      "address": "17dKfNjkUT9HGUS5hLStxMQyHD3NCF6vRY",
      "Time_step": 41.0,
      "num_txs_as_sender": 1.0,
      "num_txs_as_receiver": 1.0,
      "first_block_appeared_in": 405323.0,
      "last_block_appeared_in": 471842.0,
      "lifetime_in_blocks": 66519.0,
      "total_txs": 2.0,
      "first_sent_block": 405323.0,
      "first_received_block": 471842.0,
      "num_timesteps_appeared_in": 2.0,
      "btc_transacted_total": 0.01361188,
      "btc_transacted_min": 0.00100683,
      "btc_transacted_max": 0.01260505,
      "btc_transacted_mean": 0.00680594,
      "btc_transacted_median": 0.00680594,
      "btc_sent_total": 0.01260505,
      "btc_sent_min": 0.0,
      "btc_sent_max": 0.01260505,
      "btc_sent_mean": 0.006302525,
      "btc_sent_median": 0.006302525,
      "btc_received_total": 0.00100683,
      "btc_received_min": 0.0,
      "btc_received_max": 0.00100683,
      "btc_received_mean": 0.000503415,
      "btc_received_median": 0.000503415,
      "fees_total": 0.00621374,
      "fees_min": 0.0005731,
      "fees_max": 0.00564064,
      "fees_mean": 0.00310687,
      "fees_median": 0.00310687,
      "fees_as_share_total": 0.000573731174,
      "fees_as_share_min": 6.311739538e-07,
      "fees_as_share_max": 0.0005731,
      "fees_as_share_mean": 0.000286865587,
      "fees_as_share_median": 0.000286865587,
      "blocks_btwn_txs_total": 66519.0,
      "blocks_btwn_txs_min": 66519.0,
      "blocks_btwn_txs_max": 66519.0,
      "blocks_btwn_txs_mean": 66519.0,
      "blocks_btwn_txs_median": 66519.0,
      "blocks_btwn_input_txs_total": 0.0,
      "blocks_btwn_input_txs_min": 0.0,
      "blocks_btwn_input_txs_max": 0.0,
      "blocks_btwn_input_txs_mean": 0.0,
      "blocks_btwn_input_txs_median": 0.0,
      "blocks_btwn_output_txs_total": 0.0,
      "blocks_btwn_output_txs_min": 0.0,
      "blocks_btwn_output_txs_max": 0.0,
      "blocks_btwn_output_txs_mean": 0.0,
      "blocks_btwn_output_txs_median": 0.0,
      "num_addr_transacted_multiple": 0.0,
      "transacted_w_address_total": 2.0,
      "transacted_w_address_min": 1.0,
      "transacted_w_address_max": 1.0,
      "transacted_w_address_mean": 1.0,
      "transacted_w_address_median": 1.0,
      "node": "17dKfNjkUT9HGUS5hLStxMQyHD3NCF6vRY",
      "anomaly_score": 78.98388671875
    }
  },
\end{lstlisting}
\subsubsection{GCN and LLM Disagree on Fraud Status}
\begin{lstlisting}
  "15qA2jzinwu4DXx3TusH9J89VF32GJ66bw": {
    "top_features": [
      [
        "blocks_btwn_txs_mean",
        0.9909999966621399
      ],
      [
        "degree",
        0.9909999966621399
      ],
      [
        "btc_transacted_median",
        0.9909999966621399
      ],
      [
        "transacted_w_address_mean",
        0.0
      ],
      [
        "num_addr_transacted_multiple",
        0.0
      ]
    ],
    "explainer_scores": {
      "blocks_btwn_txs_mean": {
        "graphlime": 0.9909999966621399
      },
      "degree": {
        "graphlime": 0.9909999966621399
      },
      "btc_transacted_median": {
        "graphlime": 0.9909999966621399
      },
      "transacted_w_address_mean": {
        "graphlime": 0.0
      },
      "num_addr_transacted_multiple": {
        "graphlime": 0.0
      }
    },
    "llm_output_structured": {
      "samples": [
        {
          "explanation": "The analysis indicates that the node has a very low transaction volume and has only received a small amount of Bitcoin. The high anomaly score suggests unusual behavior, but the lack of multiple transactions or interactions with other addresses indicates it may not be involved in any fraudulent activity.",
          "is_fraud": false,
          "fraud_type": null,
          "confidence": 0.85,
          "evidence": {
            "features": [
              "blocks_btwn_txs_mean",
              "degree",
              "btc_transacted_median"
            ],
            "behaviors": [
              "single transaction activity",
              "low transaction volume",
              "no outgoing transactions"
            ]
          }
        },
        {
          "explanation": "The analysis indicates that the wallet has a very low transaction volume and has only received a small amount of Bitcoin without sending any. The anomaly score is significantly high, suggesting unusual behavior for a wallet with such limited activity. However, the lack of multiple transactions or interactions with other addresses reduces the likelihood of it being involved in fraudulent activities.",
          "is_fraud": false,
          "fraud_type": null,
          "confidence": 0.85,
          "evidence": {
            "features": [
              "btc_transacted_total",
              "num_txs_as_sender",
              "degree"
            ],
            "behaviors": [
              "single transaction activity",
              "no outgoing transactions",
              "low transaction volume"
            ]
          }
        },
        {
          "explanation": "The analysis indicates that the wallet has only one transaction as a receiver and no transactions as a sender, which is atypical behavior for legitimate wallets. The high anomaly score suggests that this wallet may be involved in suspicious activities, potentially indicating a giveaway scam.",
          "is_fraud": true,
          "fraud_type": "giveaway scams",
          "confidence": 0.85,
          "evidence": {
            "features": [
              "blocks_btwn_txs_mean",
              "degree",
              "btc_transacted_median"
            ],
            "behaviors": [
              "Only one transaction recorded",
              "No outgoing transactions",
              "High anomaly score"
            ]
          }
        },
        {
          "explanation": "The analysis indicates a high anomaly score and unusual transaction patterns, with the wallet having only one transaction as a receiver and no outgoing transactions. This behavior is characteristic of potential money laundering activities, where funds are received but not sent out, suggesting an attempt to obscure the source of the funds.",
          "is_fraud": true,
          "fraud_type": "money laundering",
          "confidence": 0.85,
          "evidence": {
            "features": [
              "anomaly_score",
              "btc_transacted_total",
              "degree"
            ],
            "behaviors": [
              "single incoming transaction",
              "no outgoing transactions",
              "high anomaly score"
            ]
          }
        },
        {
          "explanation": "The analysis indicates that the node has a very low transaction volume and has only received a minimal amount of Bitcoin. The high anomaly score suggests unusual behavior, but the lack of multiple transactions or interactions with other addresses indicates that it may not be involved in fraudulent activities.",
          "is_fraud": false,
          "fraud_type": null,
          "confidence": 0.85,
          "evidence": {
            "features": [
              "blocks_btwn_txs_mean",
              "degree",
              "btc_transacted_median"
            ],
            "behaviors": [
              "single transaction activity",
              "low transaction volume",
              "no outgoing transactions"
            ]
          }
        }
      ],
      "consensus": {
        "node_id": "15qA2jzinwu4DXx3TusH9J89VF32GJ66bw",
        "is_fraud": false,
        "fraud_type": null,
        "agreement_rate": 0.6,
        "fraud_type_agreement": null,
        "avg_confidence": 0.85
      },
      "agreement_rate": 0.6,
      "fraud_type_agreement": null,
      "avg_confidence": 0.85,
      "avg_latency": 4.857560825347901,
      "total_cost_usd": 0.0012567,
      "total_input_tokens": 5370,
      "total_output_tokens": 752,
    "node_data": {
      "degree": 3,
      "in_degree": 0,
      "out_degree": 0,
      "node_type": "unknown",
      "type": "wallet",
      "address": "15qA2jzinwu4DXx3TusH9J89VF32GJ66bw",
      "Time_step": 43.0,
      "num_txs_as_sender": 0.0,
      "num_txs_as_receiver": 1.0,
      "first_block_appeared_in": 475876.0,
      "last_block_appeared_in": 475876.0,
      "lifetime_in_blocks": 0.0,
      "total_txs": 1.0,
      "first_sent_block": 0.0,
      "first_received_block": 475876.0,
      "num_timesteps_appeared_in": 1.0,
      "btc_transacted_total": 0.0001,
      "btc_transacted_min": 0.0001,
      "btc_transacted_max": 0.0001,
      "btc_transacted_mean": 0.0001,
      "btc_transacted_median": 0.0001,
      "btc_sent_total": 0.0,
      "btc_sent_min": 0.0,
      "btc_sent_max": 0.0,
      "btc_sent_mean": 0.0,
      "btc_sent_median": 0.0,
      "btc_received_total": 0.0001,
      "btc_received_min": 0.0001,
      "btc_received_max": 0.0001,
      "btc_received_mean": 0.0001,
      "btc_received_median": 0.0001,
      "fees_total": 0.0001,
      "fees_min": 0.0001,
      "fees_max": 0.0001,
      "fees_mean": 0.0001,
      "fees_median": 0.0001,
      "fees_as_share_total": 1.7059024223814403e-08,
      "fees_as_share_min": 1.7059024223814403e-08,
      "fees_as_share_max": 1.7059024223814403e-08,
      "fees_as_share_mean": 1.7059024223814403e-08,
      "fees_as_share_median": 1.7059024223814403e-08,
      "blocks_btwn_txs_total": 0.0,
      "blocks_btwn_txs_min": 0.0,
      "blocks_btwn_txs_max": 0.0,
      "blocks_btwn_txs_mean": 0.0,
      "blocks_btwn_txs_median": 0.0,
      "blocks_btwn_input_txs_total": 0.0,
      "blocks_btwn_input_txs_min": 0.0,
      "blocks_btwn_input_txs_max": 0.0,
      "blocks_btwn_input_txs_mean": 0.0,
      "blocks_btwn_input_txs_median": 0.0,
      "blocks_btwn_output_txs_total": 0.0,
      "blocks_btwn_output_txs_min": 0.0,
      "blocks_btwn_output_txs_max": 0.0,
      "blocks_btwn_output_txs_mean": 0.0,
      "blocks_btwn_output_txs_median": 0.0,
      "num_addr_transacted_multiple": 0.0,
      "transacted_w_address_total": 2.0,
      "transacted_w_address_min": 1.0,
      "transacted_w_address_max": 1.0,
      "transacted_w_address_mean": 1.0,
      "transacted_w_address_median": 1.0,
      "node": "15qA2jzinwu4DXx3TusH9J89VF32GJ66bw",
      "anomaly_score": 77.05048370361328
    }
  },
\end{lstlisting}
\subsection{Dashboard Examples}\label{sec:DashExamples}

\begin{figure}[H]
    \centering
    \includegraphics[width=1.0\linewidth]{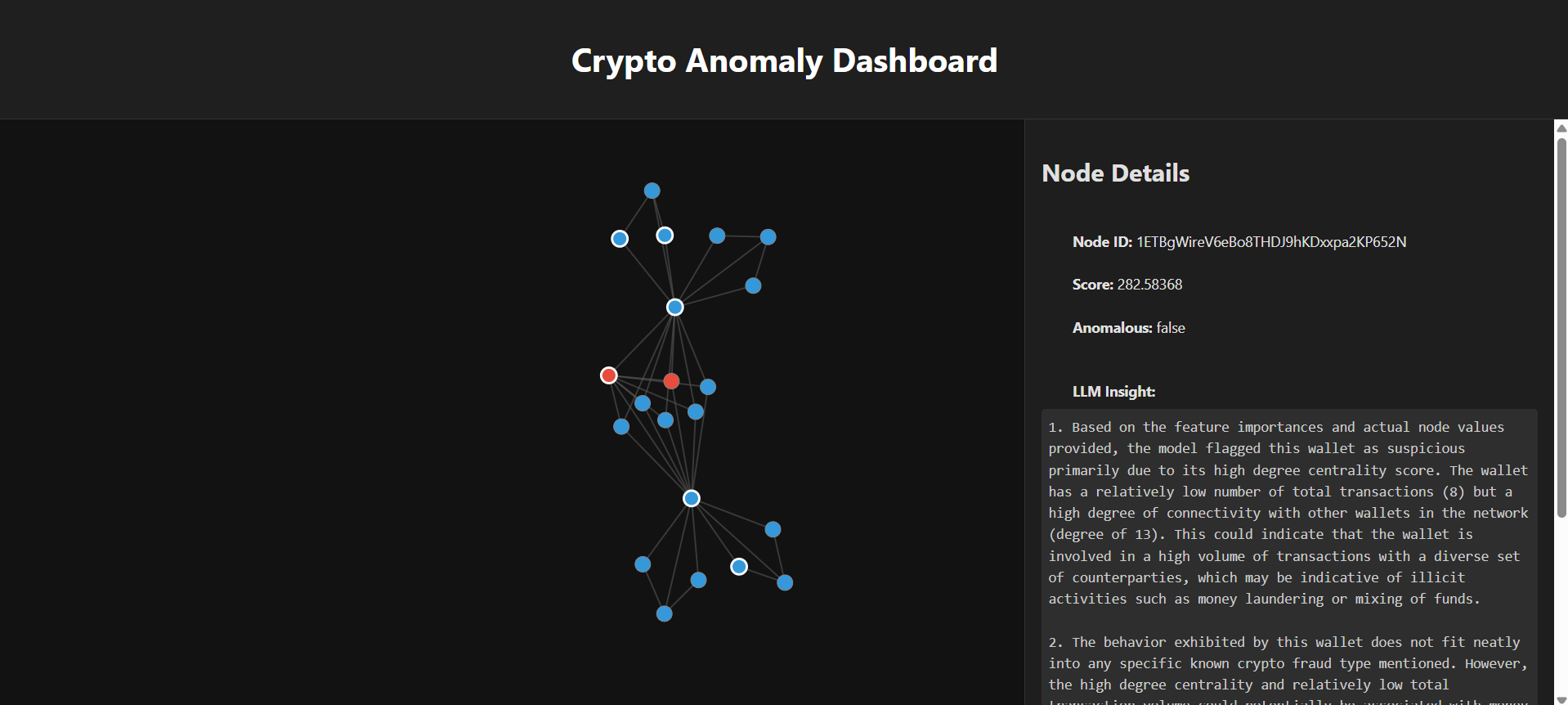} 
    \caption{A Sample of the Dashboard with the LLM Explanation and k=1}
    \label{fig:DashExample2}
\end{figure}

\begin{figure}[H]
    \centering
    \includegraphics[width=1.0\linewidth]{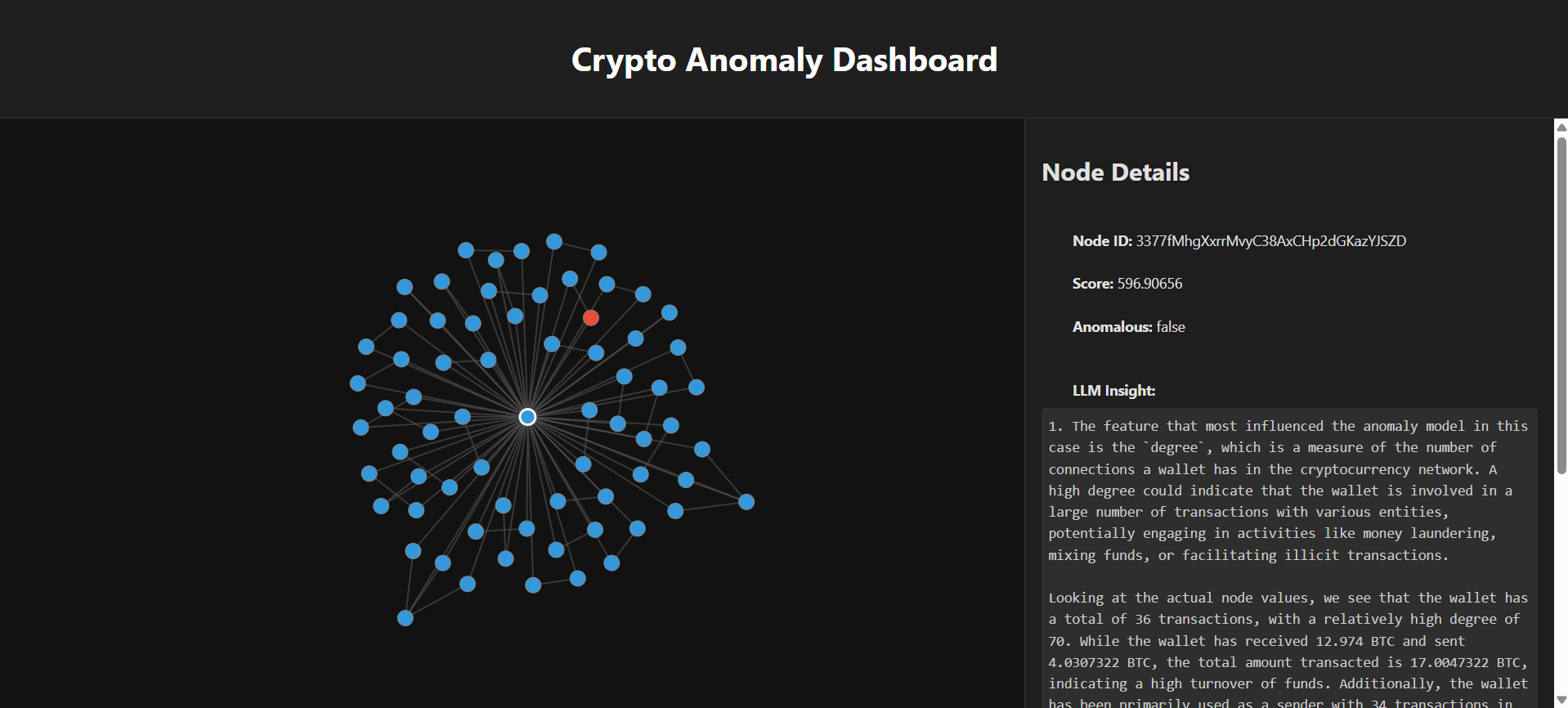} 
    \caption{A Sample of the Dashboard with the LLM Explanation and k=1}
    \label{fig:DashExample3}
\end{figure}

\subsection{Code Availability}\label{sec:CodeAvailabilty}
The codebase, including the anomaly detection model, GraphLIME explanations, and the dashboard interface, is available at: \url{https://github.com/awatson246/crypto-anomaly-detection-policy}
\end{document}